\documentclass[12pt]{article}
\title{\bf Potential regime for heavy quarks dynamics and Lorentz nature of
confinement}
\author{Yu.S.Kalashnikova\thanks{e-mail: yulia@vxitep.itep.ru},
A.V.Nefediev\thanks{e-mail: nefediev@vxitep.itep.ru}}
\date{\it Institute of Theoretical and Experimental Physics, 117259, Moscow,
Russia}
\newcommand{\ds}{\displaystyle}
\newcommand{\be}{\begin{equation}}
\newcommand{\ee}{\end{equation}}
\newcommand{\vx}{\vec{x}}
\newcommand{\vy}{\vec{y}}
\newcommand{\vz}{\vec{z}}
\newcommand{\vk}{\vec{k}}
\newcommand{\vp}{\vec{p}}
\newcommand{\vn}{\vec{n}}
\newcommand{\vg}{\vec{\gamma}}
\newcommand{\ld}{\lambda}
\newcommand{\cor}{D(\tau,\ld)}
\newcommand{\lm}{\mathop{\approx}\limits_{r\gg T_g}}
\begin{document}
\maketitle

\begin{abstract}
Propagation of the heavy quark in the field of a static antiquark source
is studied in the framework of effective Dirac equation. The model of QCD
vacuum is described by bilocal gluonic correlators. In the heavy quark
limit the effective interaction is reduced to the potential one with
5/6 Lorentz scalar and 1/6 Lorentz vector linear confinement, while
spin--orbit term is in agreement with Eichten--Feinberg--Gromes results.
New spin--independent corrections to the leading confining regime are
identified, which arise due to the nonlocality of the interaction in time
direction and quark Zitterbewegung.
\end{abstract}

A lot of evidence exists that the non-abelian nature of Yang-Mills QCD leads
to the confinement of colour charges. Apart from purely theoretical
considerations, the main bulk of data on confinement comes from the lattice
QCD simulations and from the phenomenology of hadronic spectra. The lattice
calculations firmly establish the linear rising force between two static
colour sources, while the hadronic masses are most successfully described with
the effective $q\bar q$ potential which is the sum of linear and Coulomb
forces. Complementary to these facts is the idea that QCD at large distances
is a string theory, and linear potential between heavy constituents is a
manifestation of the string--type dynamics.

In general, the dynamics governed by QCD should be nonlocal; nevertheless, it
is natural to assume that in the heavy quark limit it is reduced to the
nonrelativistic local potential acting between quark and antiquark 
which is supplied by
subleading $O(1/m^2)$ corrections. Among these corrections an important role
is played by spin-dependent forces which define the spin splittings of heavy
quarkonia and serve as a testing ground for various theoretical approaches.

The most consistent derivation of the spin-dependent potentials was performed
in the framework of the Wilson loop approach \cite{1, 2}, where the potentials
were expressed in terms of expectation values of gluonic field insertions into
a Wilson loop corresponding to the propagation of $q\bar q$ pair, and
fundamental relations between static and spin-dependent potentials were
established \cite{2}.

One of the challenging problems is the long range part of the spin-orbit force
\be
V_{SO}(r)=\left(
\frac{\vec{\sigma}_q\vec{l}_q}{4m_q^2}-
\frac{\vec{\sigma}_{\bar{q}}\vec{l}_{\bar{q}}}{4m_{\bar{q}}^2}\right)
\left(\frac1r\frac{\partial\varepsilon}{\partial r}+\frac2r
\frac{\partial V_1}{\partial r}\right)-\frac{1}{2m_qm_{\bar{q}}}
\left(\vec{\sigma}_q\vec{l}_q-
\vec{\sigma}_{\bar{q}}\vec{l}_{\bar{q}}\right)\frac1r\frac{\partial V_2}
{\partial r},
\ee
which is sensitive to the Lorentz nature of the static confining interaction
$\varepsilon (r)$. If it is a local Lorentz scalar potential, one has
\be
V_1=-\varepsilon\;\quad V_2=0,
\label{2}
\ee 
while for the time component of the Lorentz vector it is 
\be
V_1=0\;\quad V_2=\varepsilon.
\ee 

Phenomenological analysis of heavy quarkonia spectra clearly
prefers possibility (\ref{2}), but it was shown within 
Vacuum Background Correlators Method [3-6]
and in the framework of the Coulomb gauge QCD Hamiltonian 
\cite{7} that relation (\ref{2}) is respected at large distances 
without {\it ad hoc} assumption
of scalar confinement. It was demonstrated in [3-6] that 
nonrelativistic reduction
of the long range interaction is not the whole story, and additional
contributions to the spin--orbit force exist, which are due to the nonlocality
of the QCD generated interaction.

The aim of the present paper is to study systematically the nonlocality 
corrections to the effective long range interaction for heavy quark. We have
found 
a new constraint on the 
parameters of the interaction, which appears to be crucial for the 
selfconsistent potential--type dynamics of heavy quarks. We adopt the
approach suggested recently in \cite{8} and, in a more simple version, in
\cite{9}, that allows, as a byproduct, to establish explicitly the 
Lorentz nature of confinement. Our derivation is restricted to the case 
of a heavy quark propagating in the field of an infinitely heavy
antiquark colour source, and is only the first step towards defining the full
dynamics of heavy quarkonia.

The starting point of approach \cite{8} is the Green function 
$S_{q\bar{Q}}$ for the $q\bar{Q}$ system, written in the Euclidean space as
\be
S_{q\bar Q}(x,y)=\frac{1}{N_C}\int
D{\psi}D{\psi^+}DA_{\mu}
\exp{\left\{-\frac14\int 
d^4x
F_{\mu\nu}^{a2} 
-\int
d^4x
{\psi^+}(-i\hat \partial -im -\hat A)\psi \right\}}\times
\label{4}
\ee
$$
\times\psi^+(x) S_{\bar Q} (x,y)\psi(y),
$$
where $S_{\bar Q} (x,y)$ is the propagator of the static antiquark placed at 
the origin. To consider the one-body limit it is convenient to choose the 
modified Fock--Schwinger gauge \cite{10}
\be
A_4(x_4,\vec{0})=0,\quad\vec{x}\vec{A}(x_4,\vec{x})=0,
\label{5}
\ee
in which $S_{\bar Q} (x,y)$ is simply
\be
S_{\bar Q}(x,y)= i\frac{1-\gamma_4}{2} \theta (x_4-y_4) e^{-M(x_4-y_4)}+
i\frac{1+\gamma_4}{2}\theta(y_4-x_4)e^{-M(y_4-x_4)}.
\label{6}
\ee

Integration over gluonic field $A_{\mu}$ in (\ref{4}) can be performed with
the result
\be
S_{q\bar Q}(x,y)=\frac{1}{N_C}\int
D{\psi}D{\psi^+}
\exp{\left\{-\int d^4x L_{eff}(\psi,\psi^+)\right\}}
\psi^+(x) S_{\bar Q} (x,y)\psi(y),
\ee
where $L_{eff}(\psi,\psi^+)$ is the effective quark Lagrangian:
$$
\int d^4x L_{eff}(\psi,\psi^+)=\int d^4x\psi^+_{\alpha}(x)
(-i\hat\partial -im)\psi^{\alpha}(x)+ 
\int d^4x\psi^+_{\alpha}(x)\gamma_{\mu}\psi^{\beta}(x)
<{A_{\mu}}^{\alpha}_{\beta}>+
$$
\be
+\frac{1}{2}\int d^4x_1d^4x_2
\psi^+_{\alpha_1}(x_1)\gamma_{\mu_1}\psi^{\beta_1}(x_1)
\psi^+_{\alpha_2}(x_2)\gamma_{\mu_2}\psi^{\beta_2}(x_2)
<{A_{\mu_1}}_{\beta_1}^{\alpha_1}(x_1){A_{\mu_2}}_{\beta_2}^{\alpha_2}(x_2)>+
\ldots,
\label{8}
\ee
where all\hspace*{0.2cm} $\alpha$-s \hspace*{0.2cm}and \hspace*{0.2cm}$\beta$-s are 
\hspace*{0.2cm}fundamental colour indeces, and 
the irreducable correlators 
$<{A_{\mu_1}}_{\beta_1}^{\alpha_1}(x_1)\ldots
{A_{\mu_n}}_{\beta_n}^{\alpha_n}(x_n)>$ of all orders should enter. The 
first one, $<{A_{\mu}}^{\alpha}_{\beta}>$, vanishes due to the gauge and
Lorentz invariances, and in what follows we keep only bilocal correlator
$<{A_{\mu}}_{\beta}^{\alpha}(x){A_{\nu}}_{\delta}^{\gamma}(y)>\equiv 
{{\cal K}_{\mu\nu}}^{\alpha\gamma}_{\beta\delta}(x,y)$ and disregard the 
contributions of higher correlators.

Using gauge invariance of the vacuum one has
\be
{{\cal K}_{\mu\nu}}^{\alpha\gamma}_{\beta\delta}(x,y)=
(\lambda_a)_{\beta}^{\alpha}(\lambda_b)_{\delta}^{\gamma}{\cal K}_{\mu\nu}^{ab}(x,y)=
2(\lambda_a)_{\beta}^{\alpha}(\lambda_a)_{\delta}^{\gamma}K_{\mu\nu}(x,y),
\ee
$$
K_{\mu\nu}(x,y)=\frac{1}{2(N_C^2-1)}{\cal K}_{\mu\nu}^{aa}(x,y),
$$
and, because of the relation 
$(\lambda_a)_{\beta}^{\alpha}(\lambda_a)_{\delta}^{\gamma}=\frac12
\delta_{\delta}^{\alpha}\delta_{\beta}^{\gamma}-\frac{1}{2N_C}
\delta_{\beta}^{\alpha}\delta_{\delta}^{\gamma}$,
expression (\ref{8}) takes the form
$$
\int d^4x L_{eff}(\psi,\psi^+)=\int d^4x\psi^+_{\alpha}(x)
(-i\hat\partial -im)\psi^{\alpha}(x)+ 
$$
\be
+\frac{1}{2}\int d^4xd^4y
\psi^+_{\alpha}(x)\gamma_{\mu}\psi^{\beta}(x)
\psi^+_{\beta}(y)\gamma_{\nu}\psi^{\alpha}(y)
K_{\mu\nu}(x,y)
\ee
in the limit $N_C\to\infty$, yielding the Schwinger--Dyson equation
\be
(-i\hat{\partial}_x-im)S(x,y)-i\int d^4zM(x,z)S(z,y)=\delta^{(4)}(x-y)
\label{11}
\ee
with the self--energy part $M(x,z)$ given by
\be
-iM(x,z)=K_{\mu\nu}(x,z)\gamma_{\mu}S(x,z)\gamma_{\nu},
\label{12}
\ee
where $S(x,y)=\frac{1}{N_C}<\psi^{\beta}(x)\psi^+_{\beta}(y)>$ is the colour
trace of the quark Green function. As in gauge (\ref{5}) Green 
function
(\ref{6}) of the static source is unity in the colour space, quantity $S(x,y)$
completely defines propagation of the colourless $q\bar Q$ object. 

Gauge condition (\ref{5}) can be rewritten as 
\be
A^a_4(x_4,\vec x)=\int_0^1 
F^a_{i4}(x_4,\alpha\vec{x})
d\alpha,\quad\hspace*{2cm}
\ee
\be
A^a_i(x_4,\vec x)=\int_0^1 
\alpha x_k F^a_{ki}(x_4,\alpha\vec{x})
d\alpha,\quad i=1,2,3,
\ee
so the average $K_{\mu\nu}$ can be expressed in terms of field strength 
correlator $<F^a_{\mu\nu}(x)F^b_{\lambda\rho}(y)>$, for which we use the
parametrization \cite{3,4}:
\be
<F^a_{\mu\nu}(x)F^b_{\lambda\rho}(y)>=\frac{\delta^{ab}}{N_C^2-1}D(x-y)
(\delta_{\mu\lambda}\delta_{\nu\rho}-\delta_{\mu\rho}\delta_{\nu\lambda})+
\Delta^{(1)},
\label{15}
\ee
where the second term $\Delta^{(1)}$ is a full derivative and does not 
contribute to the confinement. As we are interested only in long range force,
we consider only the term proportional to $D(x-y)$ in (\ref{15}), which,
in contrast to $\Delta^{(1)}$, contributes to 
the area law with the string tension 
\be
\sigma=2
\int_0^\infty d\tau\int_0^\infty d\lambda D(\tau,\lambda).
\label{16}
\ee

Function $D(u_4,|\vec{u}|)$ is actually a function of $u_4^2+\vec{u}^2$ due
to Lorentz invariance, but in our apparently non-invariant treatment we
keep dependences on $|\vec{u}|$ and $|u_4|$ separately as in (\ref{16}).

Finally, for average $K_{\mu\nu}(x,y)$ one has $(\tau=x_4-y_4)$:
\be
\begin{array}{l}
K_{44}(\tau,\vx,\vy)=(\vx\vy)\int_0^1d\alpha\int_0^1 d\beta 
D(\tau,|\alpha\vx-\beta\vy|),\\
{}\\
K_{i4}(\tau,\vx,\vy)=K_{4i}(\tau,\vx,\vy)=0,\\
{}\\
K_{ik}(\tau,\vx,\vy)=((\vx\vy)\delta_{ik}-y_ix_k)
\int_0^1\alpha d\alpha\int_0^1 \beta d\beta 
D(\tau,|\alpha\vx-\beta\vy|).
\end{array}
\ee
 
The system of equations (\ref{11}), (\ref{12}) may be rewritten in terms 
of wave functions as
\be
(-i\hat{\partial}_x-im)\psi(x)+\int d^4z
K_{\mu\nu}(x,z)\gamma_{\mu}S(x,z)\gamma_{\nu}\psi(z)=0.
\label{18}
\ee

This equation is essentially nonlinear, since the eigenfunctions $\psi_n$
enter the spectral representation for $S(x,z)$. In the heavy quark limit
we solve it perturbatively, substituting the free Green function $S_0(x,z)$
into the self--energy part. The resulting linear equation was considered in
\cite{8,9}.

As correlators (\ref{15}) are defined in the Euclidean space, it is 
convenient to formulate the eigenvalues problem for equation (\ref{18})
in the Euclidean space too and perform the Wick rotation 
to the Minkowski one afterwards arriving at the Dirac-type equation
\be
(\vec{\alpha}\hat{\vec{p}}+\gamma_0 m+\gamma_0\hat{M})\psi=E\psi,
\label{19}
\ee
with operator $\hat{M}$ given by
$$
\hat{M}(\vx,\vz)=-i\int_0^\infty d\tau K_{\mu\nu}(\tau,\vx,\vz)
\int\frac{d^3\vk}
{(2\pi)^3}e^{i\vk(\vx-\vz)}\times\hspace*{6cm}
$$
\be
\hspace*{2.5cm}
\times\left\{
\gamma_{\mu}\frac{i\gamma_4\varepsilon+\vg\vk+im}{2\varepsilon}\gamma_{\nu}
e^{-(\varepsilon-E)\tau}+
\gamma_{\mu}\frac{-i\gamma_4\varepsilon+\vg\vk+im}{2\varepsilon}\gamma_{\nu}
e^{-(\varepsilon+E)\tau}
\right\},
\label{20}
\ee
where $\varepsilon=\varepsilon(\vk)=\sqrt{\vk^2+m^2}$ and $\gamma$-matrices
are Euclidean ones $(\gamma_{4E}=\gamma_{0M},\;\vg_{E}=-i\vg_{M})$. 

Operator $\hat{M}$ is nonlocal both in space and time, and our strategy is to
find the leading local limit, to establish the first order nonlocal 
corrections and to check if they are small. To this end we 
approximately rewrite the 
expression in the curly brackets in (\ref{20}) in the form
$$
i\gamma_{\mu}\frac{1+\gamma_4}{2}\gamma_{\nu}\left(1+\left(\varepsilon_0-
\frac{\vk^2}{2m}\right)\tau\right)+
i\gamma_{\mu}\frac{1-\gamma_4}{2}\gamma_{\nu}e^{-2m\tau}+\hspace*{3cm}
$$
\be
\hspace*{4cm}
+\gamma_{\mu}\left(\frac{\vk\vg}{2m}-\frac{i\vk^2}{4m^2}\right)\gamma_{\nu}
\left(1+e^{-2m\tau}\right),
\label{21}
\ee
introducing the quarks binding energy $\varepsilon_0=E-m$.

The leading local confining interaction 
is obtained after omitting the terms proportional to $\varepsilon_0$,
$\vg\vk$ and $\vk^2$ in (\ref{21}), and yields
$$
\hat{M}_0(\vx,\vz)=\delta^{(3)}(\vx-\vz)\int_0^\infty d\tau 
K_{\mu\nu}(\tau,\vx,\vx)
\left\{\gamma_{\mu}\frac{1+\gamma_4}{2}\gamma_{\nu}+
\gamma_{\mu}\frac{1-\gamma_4}{2}\gamma_{\nu}e^{-2m\tau}
\right\}=\hspace*{1cm}
$$
\be
\hspace*{5cm}=\delta^{(3)}(\vx-\vz)V_{conf}(\vx).
\label{22}
\ee

It is relevant now to comment on the Lorentz nature of confinement
and to clarify some related confusions. The underlying interaction is 
bilinear in vector verteces, as it is clearly seen from the effective  
Lagrangian (\ref{8}): everywhere including 
resulting expression (\ref{22}) the interaction contains  
$\gamma_{\mu}\ldots\gamma_{\nu}$ product. As it will be shown below, 
for heavy quarks this structure actually reduces to 
$\gamma_0\ldots\gamma_0$ product, and we 
agree at this point with the statement made in \cite{7}: the interaction
is time--like vector one. 
However, this almost trivial observation does not straightforwardly help
to answer another question, which is usually put in 
connection with the problem of the Lorentz structure: the effective
local confining 
interaction $\hat{M}_0(\vx,\vz)$
can be proportional either to unity or to $\gamma_0$,
and it is added either to the mass term (scalar confinement) or to the
energy term (vector confinement) in the effective Dirac equation (\ref{19}).  
The answer to this question depends on the behaviour of the function 
$K_{\mu\nu}$.

The correlator $D(u_4,|\vec{u}|)$ should decrease in all directions in the 
Euclidean space so that the string tension (\ref{16}) were finite, and the
Dirac structure of the confining potential depends on the correlation length 
$T_g$ which governs this decrease $(r=|\vx|)$:
$$
V_{conf}(r)=r(1+\gamma_4)\int_0^\infty d\tau\int_0^r d\lambda D(\tau,\lambda)
\left\{1-\frac{\ld}{r}+\left(\frac23-\frac{\ld}{r}+\frac{\ld^3}{3r^3}\right)
e^{-2m\tau}\right\}+
$$
\be
\hspace*{2cm}+r(1-\gamma_4)\int_0^\infty d\tau\int_0^r d\lambda D(\tau,\lambda)
\left\{\frac23-\frac{\ld}{r}+\frac{\ld^3}{3r^3}+\left(1-\frac{\ld}{r}\right)
e^{-2m\tau}\right\}
\label{23}
\ee
with large distance $(r\gg T_g)$ behaviour
\be
V_{conf}(r)=\left(\frac56+\frac16\gamma_4\right)\sigma r+O\left(
\frac{\sigma r}{mT_g}\right)
\label{24}
\ee
for $mT_g\gg 1$, and
\be
V_{conf}(r)=\frac53\sigma r+O\left(\sigma rmT_g\right)
\label{25}
\ee
for $mT_g\ll 1$. Both regimes were established in \cite{8,9}. In what 
follows we demonstrate that only the regime $mT_g\gg 1$ is selfconsistent.

Now we consider various corrections to the leading interaction (\ref{23}).
The terms proportional to $(\vg\vk)$ in (\ref{21}) 
are calculated by means of integration 
by parts and give
\be
\hat{M}_{\vg\vk}(\vx,\vz)=\delta^{(3)}(\vx-\vz)V_{\vg\vk}(\vx),
\ee
$$
V_{\vg\vk}(\vx)=-\frac{i}{2m}\int_0^\infty d\tau (1+e^{-2m\tau})\left\{
\gamma_{\mu}\gamma_i\gamma_{\nu}K_{\mu\nu}(\tau,\vx,\vx)\hat{p}_i-
i\gamma_{\mu}\gamma_i\gamma_{\nu}\frac{\partial}{\partial z_i}
K_{\mu\nu}(\tau,\vx,\vz)|_{\vz=\vx}
\right\}=
$$
$$
=\frac{1}{m}\int_0^\infty d\tau (1+e^{-2m\tau})\int_0^r d\ld \cor\left\{
(r-\ld)i(\vg\hat{\vp})+
\left(\frac23r-\ld+\frac{\ld^3}{3r^2}\right)i(\vg\vn)(\vn\hat{\vp})+
\right.
$$
$$
\hspace*{10cm}+\left.\left(\frac32-\frac{\ld}{r}\right)(\vg\vn)\right\},
$$
$$
\vn=\frac{\vx}{r},
\quad\hat{\vp}=-i\frac{\partial}{\partial \vx}.
$$

Problems start with the terms proportional to $\vk^2$ in (\ref{21}).
To bring these terms into local form one should integrate by parts twice:
\be
\hat{M}_{\vk^2}(\vx,\vz)=\delta^{(3)}(\vx-\vz)V_{\vk^2}(\vx),
\ee
$$
V_{\vk^2}(\vx)=\frac{1}{4m^2}\int_0^\infty d\tau \gamma_{\mu}
(1+e^{-2m\tau}+(1+\gamma_4)m\tau)\gamma_{\nu}\left\{
\frac{\partial^2}{\partial\vz^2}K_{\mu\nu}(\tau,\vx,\vz)|_{\vz=\vx}+\right.
$$
$$
\hspace*{2cm}\left.+2i\frac{\partial}{\partial z_i}
K_{\mu\nu}(\tau,\vx,\vz)|_{\vz=\vx}\hat{p_i}
-K_{\mu\nu}(\tau,\vx,\vx)\hat{\vp}^2
\right\},
$$
and only momentum dependent terms can be expressed in terms of integrals of 
the function $\cor$:
$$
V_{\vk^2}(\vx)=\frac{1}{m^2}\int_0^\infty d\tau\int_0^r d\ld \cor\left\{
(1+e^{-2m\tau}+(1+\gamma_4)m\tau)\left[
-\frac{1}{2}(r-\ld)\hat{\vp^2}+\frac{i}{2}(\vn\hat{\vp})\right]+\right.
$$
\be
+(1+e^{-2m\tau}+(1-\gamma_4)m\tau)\left[
-\frac{1}{2}\left(\frac23r-\ld+\frac{\ld^3}{3r^2}\right)\hat{\vp}^2
+\frac{i}{3}\left(1-\frac{\ld^3}{r^3}\right)(\vn\hat{\vp})\right.-
\ee
$$
\left.\left.-\frac{1}{2r}\left(\frac23-\frac{\ld}{r}+\frac{\ld^3}{3r^3}\right)
(\vec{\sigma}\hat{\vec{l}})\right]\right\}+V_{\vk^2}({\rm momentum\;
independent}),
$$
$$
\hat{l}_i=\varepsilon_{ijk}x_j\hat{p}_k.
$$

Moreover, the momentum independent contribution appears to diverge in the 
$T_g\to 0$ limit. Indeed,
$$
V_{\vk^2}({\rm momentum\;independent})=
\frac{1}{2m^2}\int_0^\infty d\tau
(1+e^{-2m\tau}+(1+\gamma_4)m\tau)\times
$$
$$
\times\int_0^r d\ld \left\{-\frac1r+\frac{2\ld}{r^2}+\frac{r}{3}\left(
1-\frac{\ld}{r}\right)\left(1+\frac{\ld}{2r}+\frac{\ld^2}{r^2}\right)
\left(\frac{2}{\ld}\frac{\partial}{\partial \ld}+\frac{\partial^2}{\partial 
\ld^2}\right)\right\}\cor+
$$
\be
+\frac{1}{2m^2}\int_0^\infty d\tau
(1+e^{-2m\tau}+(1-\gamma_4)m\tau)\times
\label{29}
\ee
$$
\times\int_0^r d\ld 
\left\{-\frac{2}{3r}\left(1-\frac{\ld}{r}\right)
\left(1-\frac{2\ld}{r}-\frac{2\ld^3}{r^3}\right)+\right.
$$
$$
\left.+\frac{2}{5}r\left(1-\frac{\ld}{r}\right)^2\left(1-\frac{\ld}{2r}
+\frac{\ld^2}{2r^2}+\frac{\ld^3}{4r^3}\right)
\left(\frac{2}{\ld}\frac{\partial}{\partial \ld}+\frac{\partial^2}{\partial 
\ld^2}\right)\right\}\cor,
$$
and for large $r$ $(r\gg T_g)$ one has the following asymptotics for expression
(\ref{29}):
\be
V_{\vk^2}({\rm momentum\;independent})=\left\{
\begin{array}{ll}
(\alpha_+(1+\gamma_4)+\alpha_-(1-\gamma_4))\frac{\ds \sigma r}{\ds mT_g},
&mT_g\gg 1\\
\beta\frac{\ds \sigma r}{\ds m^2T_g^2},& mT_g\ll 1,
\end{array}
\right.
\label{30}
\ee  
where $\alpha_+$, $\alpha_-$ and $\beta$ are some coeffitients of order unity
depending on the explicit form of function $\cor$. It is clear from 
expression (\ref{29}) and asymptotics (\ref{30}) that for the case $mT_g\gg 1$
$V_{\vk^2}$ is indeed a corection to the leading regime (\ref{23}), 
(\ref{24}). For $mT_g\ll 1$ this correction is larger than the leading 
interaction, and we conclude in such a way that the embarrasing regime 
(\ref{25}) is not potential, and, moreover, for such light quarks one 
should turn to the full system of equations (\ref{11}) and (\ref{12}).

The term proportional to the binding energy is simply  
\be
V_{\varepsilon_0}(\vx)=\varepsilon_0
\int_0^\infty d\tau \tau K_{\mu\nu}(\tau,\vx,\vx)
\left\{\gamma_{\mu}\frac{1+\gamma_4}{2}\gamma_{\nu}+\gamma_{\mu}
\frac{1-\gamma_4}{2}\gamma_{\nu}e^{-2m\tau}\right\},
\label{31}
\ee
and does not cause many additional problems, it is small comparing to the 
confinement term if $\varepsilon_0 T_g\ll 1$, as it was already found in 
\cite{9}.

The effective Schroedinger equation is obtained from Dirac equation (\ref{19})
by the standard Foldy--Wounthuysen (FW) reduction, and displays  a lot of 
pleasant features for $mT_g\gg 1$. For the upper Dirac component of the 
quark wave function the resulting Hamiltonian is
\be
H_{FW}=m+\frac{\vp^2}{2m}+\varepsilon_E(r)+
V_{LS}(r)+\varepsilon_M(r)+V_{SI}(r), 
\ee
where $\varepsilon_E(r)$ is the standard confining interaction
\be
\varepsilon_E(r)=2\int_0^\infty d\tau\int_0^r d\ld\cor (r-\ld),
\label{33}
\ee
and
\be
V_{LS}(r)=-\frac{\vec{\sigma}\vec{l}}{2m^2r}
\int_0^\infty d\tau\int_0^r d\ld\cor \left(1-\frac{2\ld}{r}\right).
\label{34}
\ee

Expressions (\ref{33}) and (\ref{34}) were obtained from equation (\ref{19})
in \cite{9}, and coincide with the ones given by the 
Vacuum Background Correlators Method [3-6]. 
As it was shown in \cite{5}, form (\ref{34}) for the spin--orbit
force is equivalent to the form obtained by Gromes \cite{2} in terms of Wilson
loop expectations. 

As for $r\gg T_g$ 
$$
\varepsilon_E=\sigma r
$$
and 
$$
V_{LS}=-\frac{\vec{\sigma}\vec{l}}{4m^2r},
$$ 
these expressions mimic scalar confinement (\ref{2}) at large distances.
Nevertheless, confining interaction (\ref{24}) is not a scalar one, at
large distances it is 5/6 scalar and 1/6 vector, and is due to the electric
correlator $K_{44}$ only as one should expect for the nonrelativistic particle.

There are three sources for the spin--orbit force (\ref{34}). One
piece comes from the FW reduction of confinement (\ref{23}), it is purely
magnetic and gives
$$
V_{LS}^{(0)}=-\frac{\vec{\sigma}\vec{l}}{3m^2r}
\int_0^\infty d\tau\int_0^r d\ld\cor \left(1-\frac{\ld^3}{r^3}\right)\lm
-\frac{\vec{\sigma}\vec{l}}{6m^2r}.
$$ 
The second one stems from the FW reduction of $\vg\vk$ term and contains both
electric and magnetic contributions yielding
$$
V_{LS}^{(1)}=\frac{\vec{\sigma}\vec{l}}{6m^2r}
\int_0^\infty d\tau\int_0^r d\ld\cor \left(1+\frac{3\ld^2}{r^2}-
\frac{\ld^3}{r^3}\right)\lm\frac{\vec{\sigma}\vec{l}}{12m^2r}.
$$ 
The third piece is from $\vk^2$ term. It is purely magnetic and its share is
$$
V_{LS}^{(2)}=-\frac{\vec{\sigma}\vec{l}}{3m^2r}
\int_0^\infty d\tau\int_0^r d\ld\cor \left(1-\frac{3\ld}{2r}+
\frac{\ld^3}{2r^3}\right)\lm-\frac{\vec{\sigma}\vec{l}}{6m^2r}.
$$ 

Magnetic confinement 
\be
\varepsilon_M(r)=2r\int_0^\infty d\tau e^{-2mT_g}
\int_0^r d\ld\cor \left(\frac23-\frac{\ld}{r}+\frac{\ld^3}{3r^3}\right),
\label{35}
\ee
behaves as $\frac{\sigma r}{mT_g}$ at large distances and it is suppressed
for large $mT_g$.

There are, of course, spin--independent corrections of the standard Darwin 
term type, which 
are $O\left(\frac{\sigma}{m^2r}\right)$ at large distances.
Another correction is given by electric momentum dependent part of $V_{\vk^2}$,
which is $O\left(\frac{\sigma T_g}{mr}\right)$, and the binding energy 
correction
(\ref{31}) contributes at the same order of magnitude. Nevertheless, the
main spin--independent correction comes from the electric part of 
$V_{\vk^2}({\rm momentum\;independent})$:
$$ 
V_{SI}(r)=
\frac{1}{m}\int_0^\infty d\tau\tau
\int_0^r d\ld \left\{-\frac1r+\frac{2\ld}{r^2}+\right.\hspace*{6cm}
$$
\be
\hspace*{3cm}\left.+\frac{r}{3}\left(
1-\frac{\ld}{r}\right)\left(1+\frac{\ld}{2r}+\frac{\ld^2}{r^2}\right)
\left(\frac{2}{\ld}\frac{\partial}{\partial \ld}+\frac{\partial^2}{\partial 
\ld^2}\right)\right\}\cor,
\label{36}
\ee
which behaves as $\frac{\sigma r}{mT_g}$ at large distances, competing with
magnetic confinement (\ref{35}).

The results of the suggested approach reproduce well--known fomulae for leading
confinement (\ref{33}) and spin--orbit (\ref{34}) forces. Our derivation does 
not appeal to the Wilson loop approach and effective QCD string at large
distances, but, as model (\ref{15}) for the QCD vacuum is compatible with
the area law, the salient features are the same. A kind of a string is
developed connecting quark and antiquark, and this string is the minimal 
string of the Vacuum Background Correlators Method [4-6], or the flux tube with 
gluonic degrees of freedom in the ground state \cite{7,11}, 
as far as we neglect
the $\Delta^{(1)}$ contributions to the field strength correlator (\ref{15}).
Nevertheless, as we deal with full Dirac Hamiltonian (\ref{19}), where both
nonlocality in time direction and Zitterbewegung are included, we are able
to disclose new important corrections (\ref{35}) and (\ref{36}).

We have demonstrated that, apart from naive nonrelativistic condition
$m\gg\sqrt{\sigma}$ and more sofisticated condition $\varepsilon_0 T_g\ll 1$
\cite{9}, another requirement, $mT_g\gg 1$, is needed for the potential-type
description of heavy quark dynamics to be valid. To what extent this 
requirement is indeed new? In the $N_C\to\infty$ limit correlators (\ref{15})
are given by the pure Yang--Mills theory, with 
single nonperturbative mass scale. 
On the other hand, model (\ref{15}) contains
two dimensional parameters, correlation length $T_g$ and $D(0)$, which is
proportional to the gluonic condensate and is related to the string tension
$\sigma$ via equation (\ref{16}). So it is not surprising that two 
phenomenological quantities, $\sqrt{\sigma}$ and $T_g^{-1}$, should be of the 
same order of magnitude, and indeed they are. The commonly accepted value
of the string tension gives $\sqrt{\sigma}\approx 0.4GeV$, while lattice
measurements \cite{12} give for the correlation length value 
$T_g^{-1}\sim 1GeV$, so the requirement $mT_g\gg 1$ does not bring drastic 
changes into our understanding of heavy quark dynamics. Much more interesting
are new corrections (\ref{35}) and (\ref{36}), which are potentially large and 
even at large distances depend on the explicit form of vacuum correlation
function $D$.
\medskip

We acknowledge useful discussions with Yu.A.Simonov.
\smallskip

This work is supported by grants 96-02-19184a and 97-02-16404 of Russian
Fundamental Research Foundation and by INTAS 94-2851 and 93-0079ext.

\end{document}